\documentclass[fleqn,10pt]{wlscirep}
\usepackage[utf8]{inputenc}
\usepackage[T1]{fontenc}
\usepackage{amssymb}
\usepackage{amsmath}
\usepackage{graphicx}
\usepackage{caption}
\usepackage{subcaption}
\usepackage{siunitx}
\usepackage{gensymb}
\usepackage{lscape}

\title{Physics-Informed Solution of The Stationary Fokker-Plank Equation for a Class of Nonlinear Dynamical Systems: An Evaluation Study}

 \author[1,*]{Hussam Alhussein}
 \author[1]{Mohammad Khasawneh}
 \author[1]{Mohammed F. Daqaq}

 \affil[1]{Laboratory of Applied Nonlinear Dynamics (LAND), Division of Engineering, New York University Abu Dhabi, Abu Dhabi, UAE}

 \affil[*]{haa385@nyu.edu}

\begin{abstract}
The Fokker-Planck (FP) equation is a linear partial differential equation which governs the temporal and spatial evolution of the probability density function (PDF) associated with the response of stochastic dynamical systems. An exact analytical solution of the FP equation is only available for a limited subset of dynamical systems. Semi-analytical methods are available for larger, yet still a small subset of systems, while traditional computational methods; e.g. Finite Elements and Finite Difference require dividing the computational domain into a grid of discrete points, which incurs significant computational costs for high-dimensional systems. Physics-informed learning offers a potentially powerful alternative to traditional computational schemes. To evaluate its potential, we present a data-free, physics-informed neural network (PINN) framework to solve the FP equation for a class of nonlinear stochastic dynamical systems.  In particular, through several examples concerning the stochastic response of the Duffing, Van der Pol, and the Duffing-Van der Pol oscillators, we assess the ability and accuracy of the PINN framework in $i)$ predicting the PDF under the combined effect of additive and multiplicative noise, $ii)$ capturing P-bifurcations of the PDF, and $iii)$ effectively treating high-dimensional systems. Through comparisons with Monte-Carlo simulations and the available literature, we show that PINN can effectively address all of the afore-described points.  We also demonstrate that the computational time associated with the PINN solution can be substantially reduced by using transfer learning.

\end{abstract}
\begin{document}

\flushbottom
\maketitle

\section{Introduction}
The dynamic behavior of numerous physical phenomena is affected by indeterministic and stochastic inputs, which play a vital role in shaping the long-time behavior of the  system. To characterize their influence on the statistical response of a dynamical system, the Fokker-Planck (FP) equation emerges as a powerful mathematical tool, which permits capturing the temporal and spatial evolution of the probability density function (PDF) of the response. The FP equation is a partial differential equation (PDE) that takes into account the drift caused by deterministic sources and the diffusion arising from stochastic fluctuations. 

Finding the solution of the FP equation is often a difficult undertaking. Exact analytical solutions are only available for a highly limited subset of dynamical systems \cite{lin1995probabilistic}. Semi-analytical methods have been developed for larger, yet still a small subset of systems using various methods, such as, stochastic averaging \cite{roberts1986stochastic}. In recent years, as computing resources have become prevalent and affordable, the use of numerical methods to approximate the solution of the FP equation has gained wider acceptance. This led to the development of computational mesh-based methods such as, Finite Elements (FE) and Finite Difference (FD) schemes \cite{langley1985finite, wojtkiewicz1997high}. These methods require dividing the computational domain into a grid of discrete points and calculating approximate solution of the discretized FP equation at these points. While these methods are effective for one-dimensional problems, they become computationally intensive for two- and higher-dimensional systems \cite{er2011new, wang2016sparse}. Furthermore, the accuracy of calculations is significantly affected by the mesh size, where finer grids result in a significant increase in computational requirements. 

In contrast, another approach to obtain the response statistics bypasses the solution of the FP equation. Statistical linearization/nonlinearization \cite{booton1954nonlinear,xu2014random, wu1987comparison}, Gaussian and non-Gaussian closure schemes \cite{iyengar1978study,wu1987comparison, crandall1980non}, and Monte-Carlo (MC) simulations fall in this category. The MC approach, used in this paper for validation, entails partitioning the numerical domain into discrete bins. Then, a Monte Carlo simulation runs the stochastic differential equation for a long time, and the PDF is estimated by counting trajectory samples that fall within their respective bins. The accuracy of the approximated PDF depends on the length of the solution trajectory, and the number of discrete bins used. It is important to note that unless the trajectory is very long and the partitioning is very fine, the resulting PDF is often very noisy to be useful. 

The recent surge of artificial intelligence in different domains has also found its way into the PDE research community. Physics-Informed Neural Networks (PINNs) have emerged as a powerful tool for solving PDEs without relying on classical methods. Their formulation, formally introduced by Raissi et al. \cite{raissi2019physics}, provided a data-driven solution (forward problem) and data-driven discovery (inverse problem) of a subset of PDEs in the context of continuous time and discrete deterministic models. PINNs have also been recently extended to solve the FP equation. In one demonstration, Zhai et al. \cite{zhai2022deep} combined PINNs with MC simulations to solve the FP equation for the PDF of several multi-dimensional stochastic systems. By incorporating both MC data and the FP differential operator in the PINN framework, the demand for large training data is significantly reduced and the learning speed is significantly increased. In another demonstration, Chen et al. \cite{chen2021solving} presented a PINN-based FP solver guided by direct numerical observations, where the Kullback-Leibler divergence theorem is used to connect data observations in the PINN framework. Recently, Guo et al. \cite{guo2022monte} used PINN together with MC simulations to solve the space-fractional advection-diffusion equations. Their results demonstrated that this approach is flexible and quite effective in tackling high-dimensional fractional PDEs.

In this work, we utilize PINN to solve the stationary FP equation governing the PDF of a variety of nonlinear stochastic dynamical systems. The distinction of our approach lies in the departure from conventional reliance on known data whether coming from MC simulations or direct numerical observations. In this regard, Xu et al. \cite{xu2019solving} presented a data-free implementation of this method. However, the method was restricted to two-dimensional problems subjected to a single source of additive noise. In this work, we illustrate the efficacy of data-free PINN in predicting the stationary joint PDF for a broader class of nonlinear dynamical systems. The contribution spans the following three domains:

\begin{itemize}
    \item We investigate the stochastic response of the nonlinear Duffing-Van Der Pol oscillator under the influence of additive white Gaussian noise. The primary goal is to assess whether the PINN method can predict qualitative topological changes (P-bifurcations) in the joint PDF. While semi-analytical methods can predict such bifurcations \cite{zakharova2010stochastic}, they often fall short of accurately capturing the phenomena when dealing with a nonlinear restoring force. In this context, PINN can emerge as an alternative to these less-reliable semi-analytical methods.

    \item  We investigate the stochastic response of the Duffing and the Van Der Pol oscillators under the combined effect of additive and multiplicative white Gaussian noise. Under the combined excitation, semi-analytical methods often fall short of predicting the response behavior. Thus, researchers often resort to numerical techniques like FEM methods. In this context, PINN offers an alternative to existing numerical methods. As it will be shown later, the meshless nature of PINNs holds the potential for a significant reduction in memory requirements.

    \item We investigate the stochastic response of the nonlinear Duffing oscillator subject to non-white noise. In such cases, the noise is modeled by passing white noise through a second-order linear filter, which increases the independent variables in the FP equation to four (four-dimensional state-space). The increase in dimensionality poses a significant challenge when employing classical numerical solvers such as FEM. In this context, PINN provides a powerful technique to solve problems that remain beyond the reach of classical methods.  
\end{itemize}
 To the authors' knowledge, none of the aforementioned problems  have been addressed in the open literature. The rest of the paper is organized as follows: Section 2 presents a brief background on the FP equation. Section 3 demonstrates the PINN framework. Section 4 presents the computational results and discussion. Section 5 presents the important conclusions. 

\section{Fokker-Planck Equation}\label{sec:FP}
Without loss of generality, the dynamic behavior of any stochastic system can be fully described by means of a set of first-order ordinary differential equations as following: 
\begin{equation}
    \frac{\text{dx}_j}{\text{dt}}=\mu_j(\mathbf{x}, t) + \sum_{r=1}^{m} g_{jr}(x, t) \eta_r(t), \quad j = 1,2, ..., n
    \label{EOM}
\end{equation}
Here, $\mathbf{x}$ represents the vector of state variables; $\eta_r(t)$ is a white Gaussian noise with zero mean and a constant cross-density $S_{rs} = E\{\eta_r, \eta_s\}$, where $r$ and $s$ range from 1 to $m$ ($m$ being the number of acting noise sources), and the operator $E\{.\}$ represents the mean value operator in the Gaussian context. Additionally, $\mu_j$ and $g_{jr}$ are the continuous deterministic functions of the state variables $\mathbf{x}$ and $t$.

The evolution of the joint PDF of the stochastic process $\mathbf{x}$ is obtained by solving the following FP Equation\footnote{More details into the derivation of the FP equation can be found in many manuscripts devoted to stochastic differential systems \cite{lin1995probabilistic, van1976stochastic, arnold1995random}} as: 
 \begin{align}    
     &\frac{\partial P(\mathbf{x}, t)}{\partial t} = \mathcal{L} P(\mathbf{x,t}) = - \sum_{j=1}^{n} \large(\kappa_j(\mathbf{x}, t) P(\mathbf{x}, t) \large)_{x_j} + \frac{1}{2} \sum_{j,k=1}^{n} \left(\kappa_{jk}(\mathbf{x}, t) P(\mathbf{x}, t) \right)_{{x_j}{x_k}}.
 \end{align}
Here, the subscripts $x_j$ denote partial derivatives; $\kappa_j$ is a vector field representing the drift function, and $\kappa_{jk}$ is a  matrix-valued function representing the diffusion coefficients. These can be expressed as 
 \begin{align}
     \kappa_j &= \mu_j(\mathbf{x}, t) + \frac{1}{2}\sum_{r,s =1}^{m} S_{rs}g_{js}(\mathbf{x}, t)\frac{\partial g_{jr}(\mathbf{x}, t)}{\partial g_j}, \nonumber \\
     \kappa_{jk} &= \sum_{r,s=1}^{m} S_{rs}g_{jr}(\mathbf{x}, t)g_{ks}(\mathbf{x}, t).
 \end{align}
 The boundary conditions on Equation (2) involves setting the value of the joint PDF to zero as $\mathbf{x} \to\infty$; that is 
 \begin{equation}
     \lim_{\mathbf{x}\to\pm\infty} P(\mathbf{x},t) = 0.
 \end{equation}
 The long-time behavior of the PDF also known as the stationary PDF of the response $P(\mathbf{x})$ is obtained by setting $\frac{\partial P(\mathbf{x})}{\partial t} = 0$.

\section{Physics-informed Learning}
\begin{figure}[t]
\centering
\includegraphics[width=0.85\textwidth]{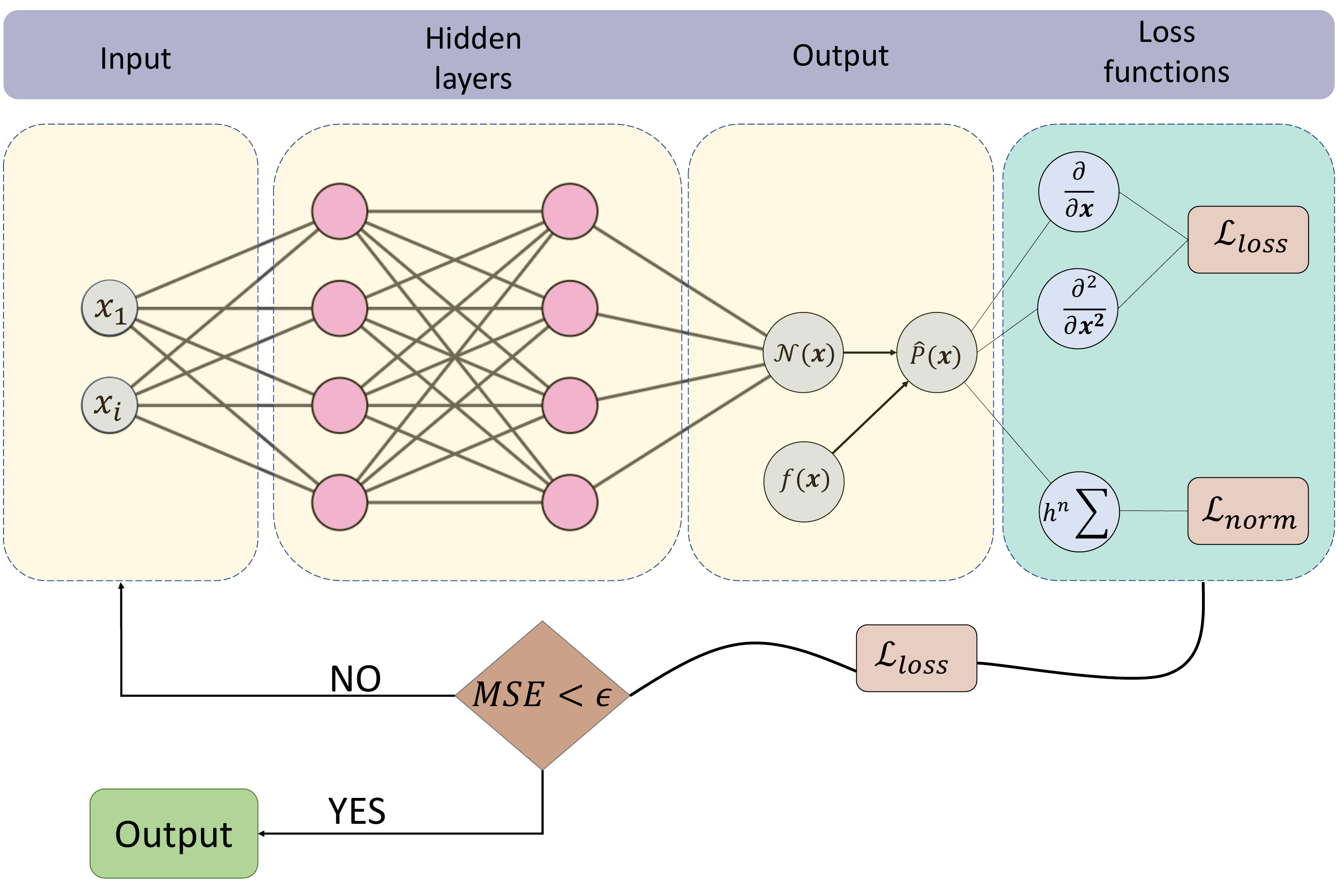}
\caption{Schematic of a physics-informed neural network (PINN). A fully-connected feed-forward neural network with space coordinates as an input to estimate $\hat{P}(\mathbf{x})$ as its output.}\label{vanillaPINN}.
\end{figure}

In the context of PINNs, a fully-connected feed-forward neural network (NN) composed of multiple-hidden layers is used to approximate the stationary solution of Equation (2). This is achieved by taking a concatenation of the state-space $\textbf{x}$ as an input and a real number as an output to approximate $P(\mathbf{x})$ as shown in Fig. \ref{vanillaPINN}. The NN is a feed-forward network meaning that each layer creates data for the next layer through the following nested transformations \cite{svozil1997introduction}:
\begin{equation}
 z^l = \sigma^l (W^l.z^{l-1} + b^l), \quad l = 1, .. , L, 
\end{equation}
where $z^0 = \mathbf{x}$ is the input, and $z^L =\mathcal{N}(\mathbf{x})$ is the output. The functions $\sigma^l$, which are called activation functions  make the network nonlinear with respect to the inputs. The learnable parameters of the NN are $W^l$ and $b^l$ denoting the weights and biases of each layer $l$, respectively.

In PINNs, solving the PDE system described by Equation (2) is converted into an unconstrained optimization problem by iteratively updating $W^l$ and $b^l$ with the goal of obtaining the solution \cite{raissi2019physics}. Based on this formulation, we have three requirements for the training as follows: 
 \begin{itemize}
     \item The output $\mathcal{N}(\mathbf{x})$ should approximately satisfy the stationary component of Equation (2). This is achieved through incorporating the FP equation into the training by introducing the following PDE loss function:
 \begin{equation}
     \mathcal{L}_{PDE} = \frac{1}{N_D}\Bigg(\sum(\mathcal{L} \mathcal{N}(\mathbf{x_i: \theta}) \Bigg)^2, \qquad i = \{1, 2, .., N_D\}
 \end{equation}
 where $\mathcal{L}_{PDE}$ is the discretized operator of $\mathcal{L}$ and represents the classic mean-square error; $\mathbf{x}_i$ are the set of uniformly distributed sample of discrete state-space variables, $N_D$ is the total number of descritization points, and $\theta$ denotes all learnable parameters of the network (e.g. weights, $W$, and biases, $b$). Note that all the required gradients with respect to input variables $\mathbf{x}$ or network parameters $\theta$ can be efficiently computed via automatic differentiation \cite{griewank2008evaluating}.  
     \item The conditions at the domain boundaries should be satisfied. Enforcing the PDF to go to zero at the infinite domain boundaries is not realistically possible. To tackle this problem, we assume a zero boundary condition on a domain that is large enough to cover the high density areas with sufficient margin \cite{zhai2022deep}. For instance, if $\partial\Gamma_D$ is the boundary of a sufficiently large domain, $D$, then we let 
  \begin{equation}
     P(\mathbf{x}) = 0, \quad \quad \mathbf{x}\in\partial\Gamma_D.
 \end{equation}
 To force the approximate solution $\mathcal{N}(\mathbf{x}: \theta)$ to satisfy the boundary conditions, we modify the network and construct the new solution as: 
\begin{equation}
    \hat{P}(\mathbf{x}: \theta) = f(\mathbf{x}) \times \mathcal{N}(\mathbf{x}:\theta),
\end{equation}
where $\hat{P}(\mathbf{x}:\theta)$ is the modified network output, and $f(\mathbf{x})$ is a function satisfying the following two conditions: 
\[
f(\mathbf{x}) = 
\begin{cases}
  f(\mathbf{x}) = 0, & \mathbf{x} \in \partial\Gamma_D, \\
  f(\mathbf{x})>0, & \text{elsewhere}.
\end{cases}
\]
If the boundary is a simple geometry, then it is possible to choose $f(\mathbf{x})$ analytically. For instance, $f(x)$ can be a assumed to take the form $f(x) = x \times (a - x)$ for a one-dimensional domain scaled over the interval $[0 \times a]$. For higher dimensional cases, a generalization to the form 
\begin{equation}
    f(\mathbf{x}) = r_0 \prod_{i=1}^{n} x_i (a_i - x_i),
\end{equation}
defined over the scaled interval  $[0 \times a_i]^n$ is necessary, where $n$ represents the number of independent variables. Here, $r_0$ is an arbitrary constant, which can be selected to ensure the peak point of the function $f(\mathbf{x})$ stays of order one. For instance, $r_0$ is chosen as 16 and 128 for, respectively, two- and four-dimensional state spaces.
     \item The total probability density remains properly normalized over the entire state-space. To this end, we add a normalization loss function to ensure that  $\int_{D}^{} P (\mathbf{x}) d\mathbf{x} = 1$. For uniformly-sampled collocation points in the domain $D$, the integral is transformed into the summation of $\hat{P}(\mathbf{x}:\theta)$ multiplied by the hypercube size. Hence, the normalization loss function in an $n$-dimensional state-space can be written as:
\begin{equation}
    \mathcal{L}_{norm} =\bigg(h^n \sum \hat{P}(\mathbf{x}:\theta)\bigg)^2 - 1,
\end{equation}
where $h$ is the spacing between any two adjacent collocation points.  

The total loss function is therefore the sum of the PDE and normalization losses as: 
\begin{equation}
    \mathcal{L}_{loss} = \mathcal{L}_{PDE} + \lambda \mathcal{L}_{norm},
\end{equation}
where $\lambda$ is a weight for scaling and balancing the loss terms. In general, we observed that $\lambda = 0.1$ ensures a proper convergence. Since the PINN solver already satisfies the boundary condition, a loss functions with respect to the boundary is not needed \cite{xu2019solving}.   Imposing a hard boundary is sufficient and is found to reduce the number of loss terms, enhance training convergence, and predictive accuracy. 
 \end{itemize} 

While satisfying the preceding requirements, the network takes the full-batch coordinates of a spatial domain, $\mathbf{x}$, and predicts the target solution, $\hat{P}(\mathbf{x})$. Because of its effectiveness, we used the \textit{tanh} activation function \cite{nwankpa2018activation}. The weights of the network are initialized with Xavier initialization \cite{glorot2010understanding},  while the biases are initialized as zero. No dropout methods were applied. For the optimization algorithm, we are looking for the best possible parameters $\theta$ that minimize the loss functions as:
\begin{equation}
    \theta^* = arg \; min_\theta \;   \mathcal{L}_{loss}(\theta).
\end{equation}
To minimize the loss function,  $\mathcal{L}_{loss}$, we combine ADAM \cite{kingma2014adam} and L-BFGS-B\cite{zhu1997algorithm} optimizers. We first apply the ADAM optimizer for gradient descent training and then employ the L-BFGS-B optimizer to fine tune the results. During the Adam-based training, the optimizer calculates the direction of the gradient at each iteration considering a full-batch size. The initial learning rate is 0.001. We set the stopping criterion of L-BFGS-B to the smallest positive normalized floating-point number represented in Python3. For all experiments, the networks are trained on a NVIDIA A600 GPU. Other MC simulations are performed on 52-core Intel Core i7-8700 CPUs.

\subsection{Optimal NN Architecture via an Illustrative Example}\label{sec:example}
In this section, we provide an illustrative example for deriving the FP equation and solving for the joint stationary PDF utilizing the PINN framework. Using this example, we conduct a systematic parametric study to determine the optimal NN architecture and optimal number of collocation points, $N_D$. 

To this end, we consider the stochastic dynamics of the widely-celebrated Duffing oscillator excited by additive white noise, $\eta_1$. The equation governing the time evolution of the dynamics can be written as:
\begin{equation}
    \Ddot{x} + 2\zeta\,\Dot{x} + \alpha\,x + \gamma\, x^3= \eta_1,
    \label{Duffing}
\end{equation}
where $\alpha$, and $\zeta>0$ are respectively the linear stiffness and damping parameters, while $\gamma>0$ is the coefficient of the nonlinear stiffness. The process $\eta_1$ is a Gaussian white noise with zero mean and density $S_{11}$. The stiffness $\alpha$ can take on both positive (monostable potential) or negative values (bistable potential). 

Equation (\ref{Duffing}) can be described in the following state-space format: 
\begin{align}
    \Dot{x}_1 &= x_2, \nonumber \\
    \Dot{x}_2 &= - 2\zeta x_2 -\alpha x_1 - \gamma x_1^3  + \eta_1,
\end{align}
The drift and diffusion coefficients follow from Equation (3) as:
\begin{align}
    &\kappa_1 = x_2, \nonumber \\
    &\kappa_2 = -\alpha x_1 -\gamma x_1^3 - 2\zeta x_2, \\
    &\kappa_{22} = S_{11}. \nonumber
\end{align}
The FP equation can then be obtained by substituting Equations (15) into Equation (2). After some adaptation one obtains: 
    \begin{equation}
    \frac{\partial P}{\partial t} = -\frac{\partial(x_2 P)}{\partial x_1} - \frac{\partial \left[ (\alpha x_1 + \gamma x_1^3 + 2\zeta x_2)) P \right]}{\partial x_2} + \frac{1}{2} \frac{\partial^2 (S_{11} P)}{\partial x_2^2}
    \label{eq:FPKexample}
\end{equation}
The stationary form of Equation (15) is one of the few examples where the FP Equation admits an exact analytical solution. We use this solution to compare with the PINN predictions in order to demonstrate several properties of the PINN solver. The solution has the following Boltzman form:
\begin{equation}
    P(x_1, x_2) = C \exp\Bigg({\frac{\zeta \alpha x_1^2}{S_{11}}}(1 - \frac{1}{2}\gamma x_1^2)\Bigg) \exp\Bigg({\frac{-\zeta}{S_{11}}x_2^2}\Bigg),
    \label{analyticalSol}
\end{equation}
where $C$ is a normalizing factor given by: 
\begin{equation}
    C^{-1} = \int_{-\infty}^{\infty}  \exp\Bigg({\frac{\zeta \alpha x_1^2}{S_{11}}}(1 - \frac{1}{2}\gamma x_1^2)\Bigg) \exp\Bigg({\frac{-\zeta}{S_{11}}x_2^2}\Bigg) dx_1 dx_2.
\end{equation}
To solve Equaton \eqref{eq:FPKexample}, we use the following numerical parameters: $\zeta= 0.2, \alpha = 1, \gamma = 1, S_{11}= 0.05$, and define the domain boundaries to extend from -2 to 2 in both $x_1$ and $x_2$ directions. We choose the number of collocation points $N_D$ = 40441 such that the spacing $h$ between any adjacent points is 0.02. The network architecture consists of 4 layers and 12 neurons. To train the network, we first use ADAM optimizer with learning rate $0.001$ for 1000 epochs, and subsequently we switch to L-BFGS-F until the total loss has converged.

\begin{figure}[t]
\centering
\includegraphics[width=0.85\textwidth]{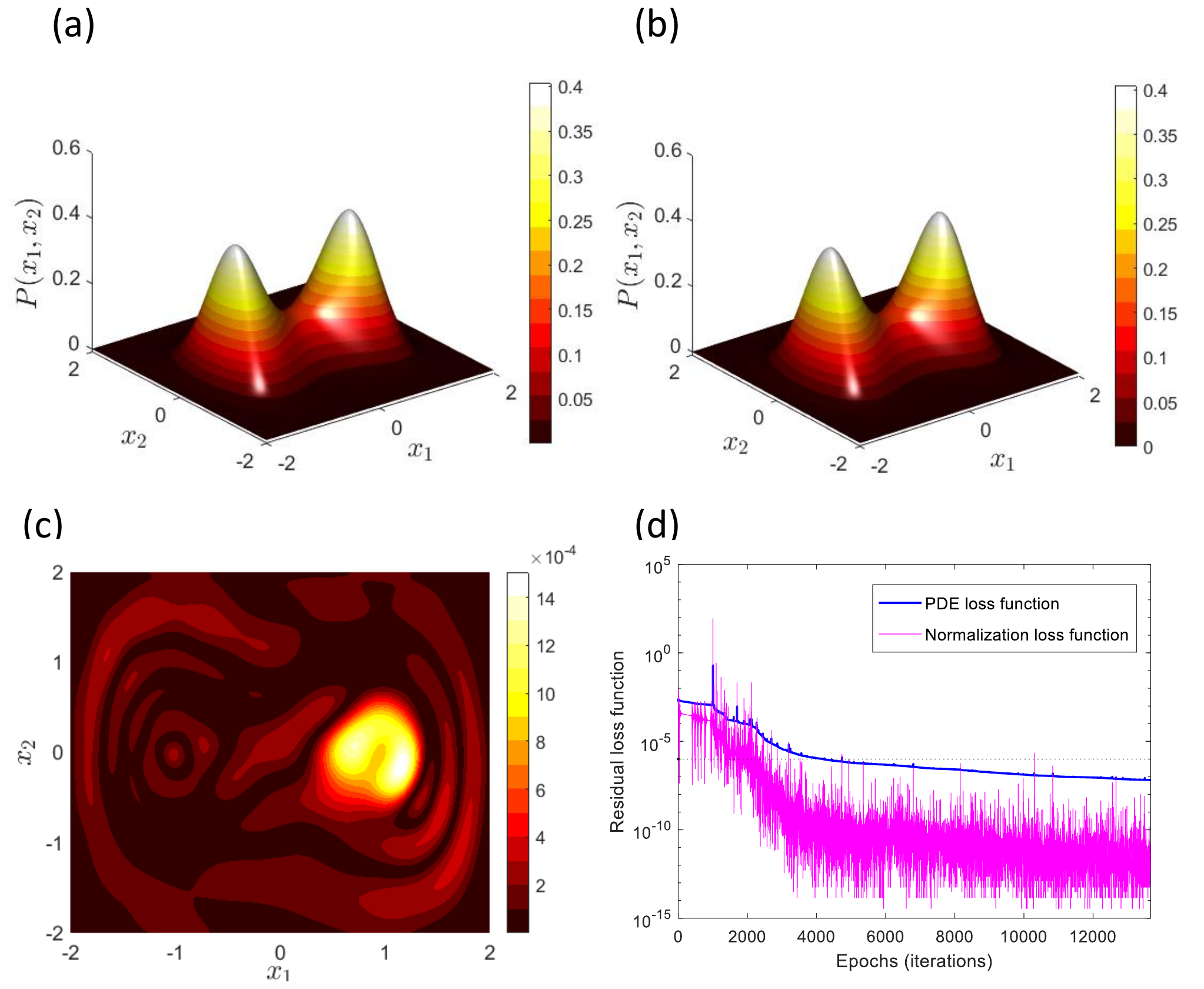}
\caption{(a) Exact analytical solution obtained from Equation \eqref{analyticalSol}. (b) Joint PDF obtained from the PINN framework. (c) Point-wise error between the analytical solution and the PINN method. (d) Convergence of the residual loss functions, namely, $\mathcal{L}_{PDE}$ and $\mathcal{L}_{norm}$.}\label{fig:Duffing1}.
\end{figure}

Results are presented in Fig. \ref{fig:Duffing1}. Upon comparing the analytical solution displayed in Fig. \ref{fig:Duffing1}(a) with the PINN prediction depicted in Fig. \ref{fig:Duffing1}(b), it becomes evident that the  two solutions are in excellent agreement. This high level of agreement is further illustrated in Fig. \ref{fig:Duffing1}(c), where the point wise error is visualized. The error remains consistently below $0.15 \times 10^{-4}$. The low error is also reflected in Fig. \ref{fig:Duffing1}(d), where the residual loss functions are plotted against the epochs (iterations) of training. It can be clearly seen that the PDE loss function, $\mathcal{L}_{PDE}$, decreases to a value below $(10^{-7})$, whereas the normalization loss, $\mathcal{L}_{norm}$, decreases to a value below $(10^{-10})$.  This clear correlation between the point wise error and the loss function terms is useful, because it can be used to check the accuracy of the prediction, when an analytical closed-form solution is not readily available. 

\begin{figure}[t]
\centering
\includegraphics[width=0.85\textwidth]{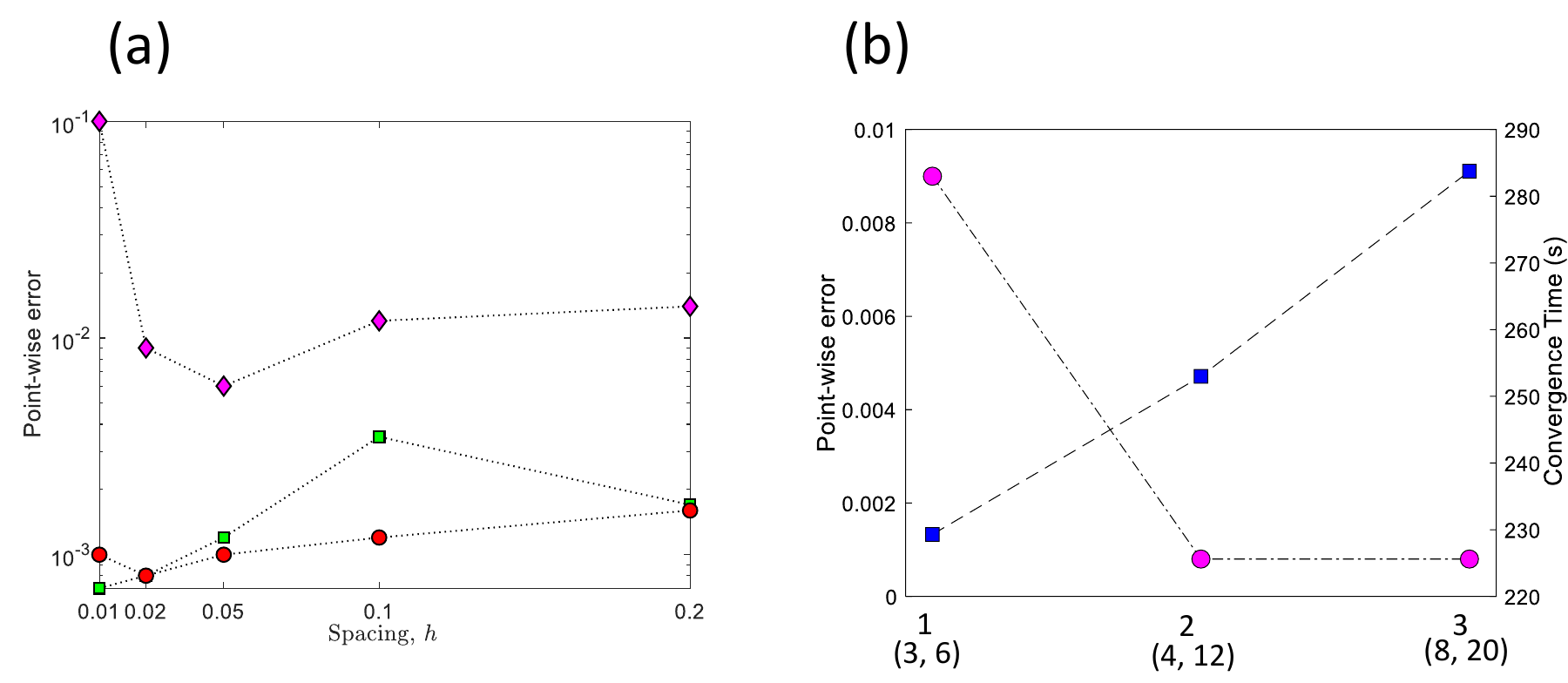}
\caption{(a) Point-wise error with varying the spacing, $h$, considering three different architectures, (diamond) 3 layers, 6 neurons,  (circles) 4 layers, 12 neurons, (squares) 8 layers, 20 neurons. (b) Comparison of the error (circles) and convergence time (squares) for the three different architectures. }\label{fig:parameteric}.
\end{figure}

The choice of an appropriate network architecture and the number of collocation points impacts the success of the training. In the previous training experiment, the selection of the network hyper parameters and the number of collocation points was achieved randomly. In the current exercise, we conduct a systematic parametric study to explore the optimal architecture. To achieve this, we conduct a comprehensive parametric study by systematically increasing both the architecture size and the number of collocation points. In particular, we compare three distinct network architectures: the first consists of 3 layers with 6 neurons, the second consists of 4 layers with 12 neurons, and the third employs 8 layers with 20 neurons. Figure \ref{fig:parameteric}(a) shows the absolute error against the spacing for the three distinct networks. When employing a relatively smaller network (diamond), the training converges, but the error remains high, particularly when the spacing between the collocation points is large. For instance, at a spacing of $h = 0.2$, the error is close to $0.014$. As the spacing is reduced to $h = 0.05$, the relative error decreases by one order of magnitude. When the interval is further reduced to $h = 0.01$, the training process fails to converge, resulting in a significantly elevated error. This indicates that a small network encounters difficulties in handling a substantial number of collocation points effectively. Upon increasing the network architecture to 4 layers and 12 neurons (circles), it is observed that the error decreases as the interval between the collocation points becomes finer. The lowest error value is observed when the interval is set to $h  = 0.02$. Architecture 3 (8 layers, 20 neurons, represented by square symbol) exhibits similar pattern to the former architecture, however the lowest error occurs when the spacing is very small $h = 0.01$. This indicates that a larger and deeper network architecture can effectively handle finer collocation point intervals.

To identify the optimal architecture in our case, we perform a comprehensive comparison of the three networks, taking into account both the error and the convergence time at a fixed spacing of $h = 0.02$, as depicted in Fig. \ref{fig:parameteric}(b). It is evident that a smaller architecture exhibits a faster convergence time at the expense of a higher error rate. When we contrast architectures 2 and 3, we find that their relative errors are quite similar, yet architecture 2 requires significantly less time to converge. As a result, we select the network architecture comprising 4 layers with 12 neurons, while maintaining a spacing interval of $h$ equal to 0.02. This choice strikes a balance between computational efficiency, as indicated by the faster convergence time, and accuracy, as reflected in the lower error.

\section{Computational Results and Discussion}
In this section, we present the results of the data-free PINN FP equation solver by exploring a diverse array of stochastic dynamical systems, each addressing an important aspect of the response behavior that has not been studied in the open literature as follows:

\begin{itemize}
    \item Prediction of P-bifurcations: This will be achieved by studying the joint PDF of the nonlinear Duffing-Van Der Pol oscillator under the influence of additive white Gaussian noise. 

    \item Combined noise sources: This will be achieved by studying the joint PDF of the Duffing and the Van Der Pol oscillators under the combined effect of additive and multiplicative white Gaussian noise. 

    \item High dimensional FP equations: This will be achieved by studying the response of the Duffing oscillator subjected to band-limited noise.  
\end{itemize}

While, in the aforementioned cases, results cannot be verified with exact analytical solutions, we qualitatively compare the PINN predictions to approximations obtained through  MC simulations. To this end, independent sets of random noise signals are generated, and when needed, they are correlated to achieve a desired partial correlation coefficient. The noise signal is then used as an input to Equation (1), which is integrated numerically using the Euler-Maruyama method to obtain $x_j$.  The joint PDF for the trajectory is estimated using the Normal Kernel Smoothing Density Function (ksdensity in Matlab \cite{MATLAB}) at equally-spaced points that cover the range of the domain \cite{}.  It is worth noting that MC simulations themselves are approximations, but they serve as a useful benchmark to demonstrate the effectiveness of the PINN framework. Whenever possible, we also draw comparisons with results from the published literature.

We would like to remark that our primary intention is not to draw comparisons between this physics-informed method and other techniques, such as FEM or MC. Instead, our focus is to demonstrate the versatility of the PINN framework in addressing a wide range of complex stochastic dynamical systems.

\subsection{Example 1: The Nonlinear Duffing-Van Der Pol Oscillator Under White Noise}
In the first illustrative example, we consider the nonlinear Duffing-Van Der Pol oscillator subject to additive white Gaussian noise. This oscillator serves as a mathematical model for a range of dynamical systems \cite{applic1, applic2}.  The equation of motion in its basic form can be written as: 
\begin{align}
    &\Dot{x}_1 = x_2, \nonumber \\
    &\Dot{x}_2 = -(-2\zeta + x^2 - x^4)\,\Dot{x} - \alpha\,x - \gamma\,x^3 + \eta_1.
\end{align}
 It is well known that the joint PDF of this system undergoes qualitative changes in its shape as either the parameters $\alpha$ and/or $\zeta$ are varied. These are commonly known as the P-bifurcations. The evolution of qualitative changes can be understood through the solution of the corresponding FP equation \cite{risken1985fokker}. Many researchers employed a first-order stochastic averaging to capture those bifurcations \cite{roberts1986stochastic}; however, it was observed that a first-order stochastic averaging is unable to capture the influence of the nonlinear stiffness terms, as these effects are removed during the averaging process. This problem can be circumvented by considering higher-order terms in the stochastic averaging, but leads to cumbersome and lengthy analytical calculations \cite{schmidt1980vibrations}. Here, we investigate whether the PINN framework can predict such bifurcations. 
 
 Following the formulation in Equation (2), the corresponding stationary FP equation can be written as: 
\begin{equation}
    0 = -\frac{\partial(x_2 P)}{\partial x_1} - \frac{\partial \left[ (-2\zeta + x_1^2 - x_1^4)x_2 P\right]}{\partial x_2} - \frac{\partial \left[ (\alpha x_1 + \gamma x_1^3)x_2 P\right]}{\partial x_2} + \frac{1}{2} \frac{\partial^2 [(S_{11} P]}{\partial x_2^2}.
    \label{eq:FP_duffingVander}
\end{equation}

To solve Equation \eqref{eq:FP_duffingVander}, the hyperparameters of PINN were derived using a comprehensive parametric study similar to the one presented in Section \ref{sec:example} leading to the following parameter values: the neural network architecture is configured with 8 layers and 20 neurons, and the spatial resolution is set to $h = 0.02$, while the learning rate is fixed to 0.001 for a total of 1000 ADAMs iterations. 

\begin{figure}[tb]
\centering
\includegraphics[width=0.85\textwidth]{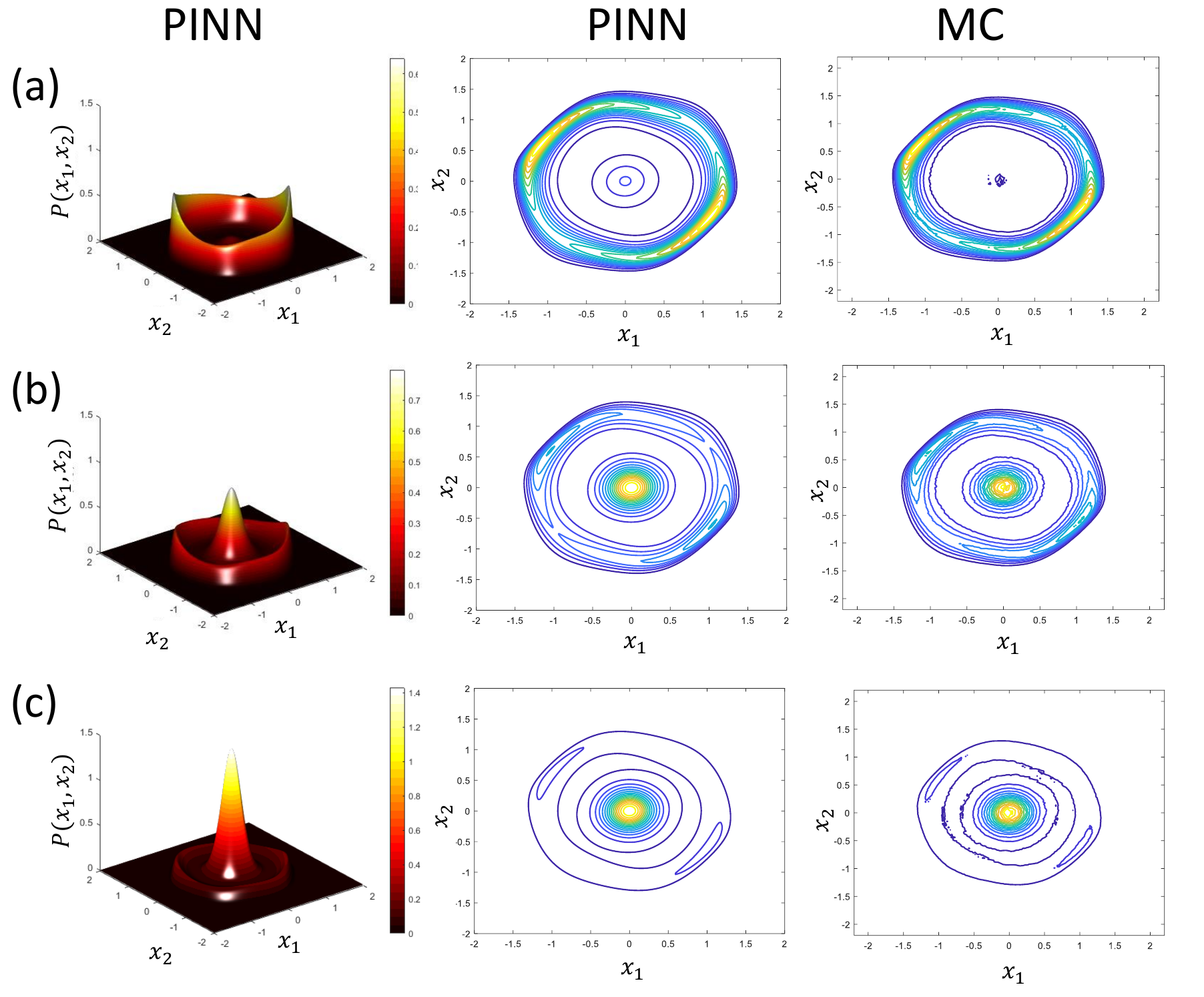}
\caption{Joint PDF of the nonlinear Duffing-Van Der Pol oscillator under white Gaussian additive noise. Results are obtained for $\alpha = 1$, $\gamma=1$, $S_{11}= 0.01$ , (a) $\zeta = 0.04$, (b) $\zeta = 0.05$ , (c) $\zeta = 0.055$.}\label{DufVanDer_c}.
\end{figure}

We examine the stationary PDF as the parameters, $\zeta$ and $\alpha$ are varied. Figure \ref{DufVanDer_c} presents the predicted PDFs when the parameter $\zeta$ is varied from $0.04$ to $0.55$, while  $\alpha$ and  $\gamma $ are kept constant at unity.  We overlay the approximated MC solutions in the third column of the figure for comparative purposes. An inspection of the figure clearly indicates that at $\zeta =0.04$, the PDF consists of a circular annular region within the phase space. This region is indicative of a stochastic attractor that mimics limit-cycle oscillations (LCO) in the deterministic sense.
At a parameter value of $\zeta = 0.05$, the circular annular region persists, and a new region of high probability emerges at the origin. This behavior suggests that the system alternates between two high-probability regimes: A stochastic LCO attractor and a trivial fixed point. At $\zeta = 0.055$, the strength of the LCO attractor diminishes and the fixed point attractor dominates. The observed topological transitions in the PDF which mimic the deterministic transition from stable limit cycles oscillations, to bi-stability, followed by elimination of one attractor is a clear indication of P-bifurcations. Upon comparing the PINN with MC simulations, we notice a significant level of agreement, even though the MC simulations are inherently coarse and less accurate. To provide further evidence of the accuracy of the PINN method, the reader can find further supporting data in reference \cite{kumar2016investigations}, where the PINN results exhibit strong agreement with FEM simulations. 

\begin{figure}[tb]
\centering
\includegraphics[width=0.85\textwidth]{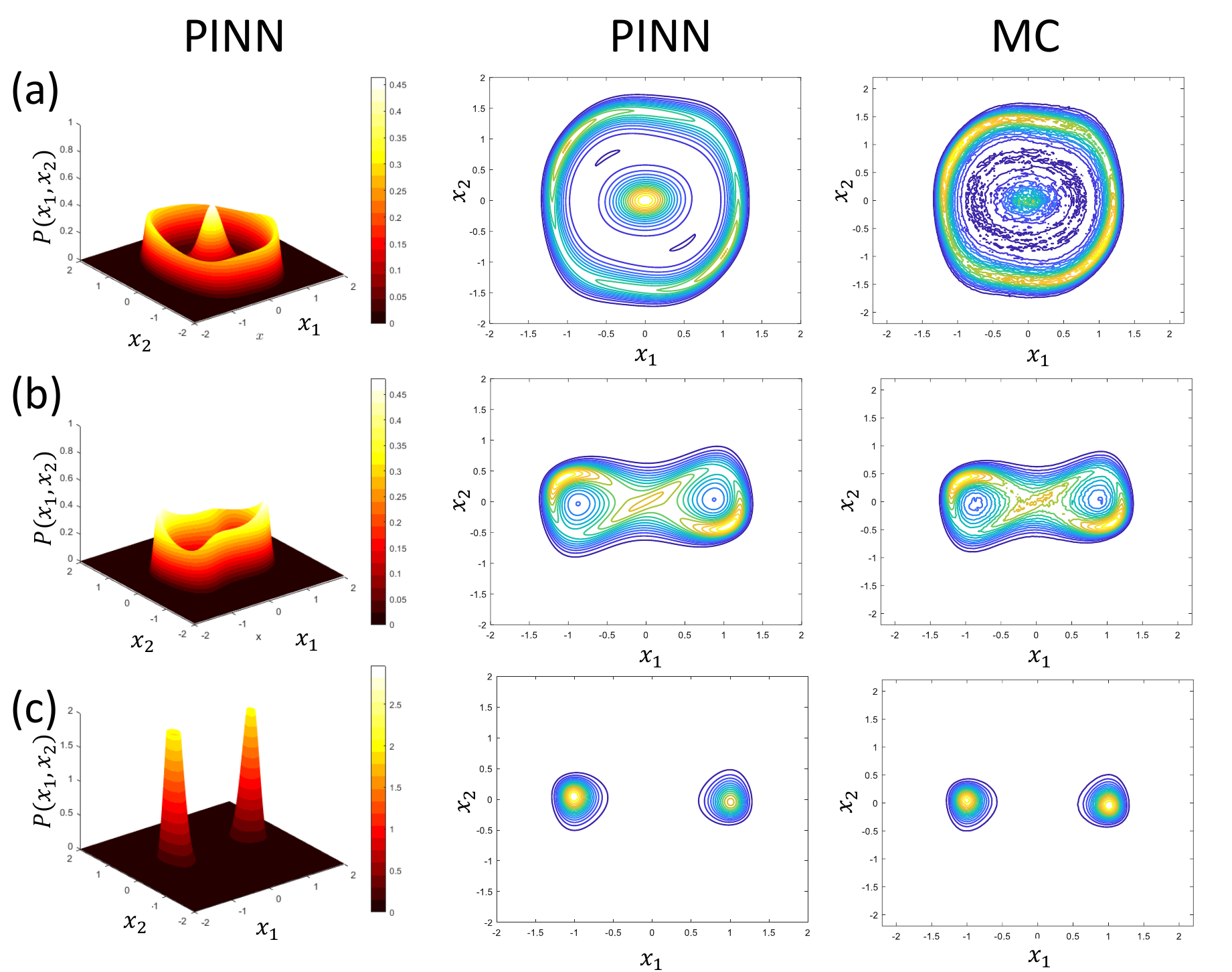}
\caption{Joint PDF of the nonlinear Duffing-Van Der Pol oscillator under white Gaussian additive noise. Results are obtained for $\zeta = 0.1$, $\gamma=1$, $S_{11}= 0.01$ , (a) $\alpha = 0.75$, (b) $\alpha = -0.75$ , (c) $\alpha = -1.0$.}\label{DufVanDer_alpha}.
\end{figure}

In Fig. \ref{DufVanDer_alpha}, we present the PDF when varying the linear stiffness $\alpha$, while keeping the linear damping constant at $\zeta = 0.1$. When $\alpha = 0.75$, the system demonstrates bistable behavior. This is evident by the existence of two stochastic attractors: a fixed point located at the origin and an LCO around it. The LCO is not symmetric about either $x_1$ or $x_2$. For $\alpha = -0.75$, a change in the system's behavior occurs. The previously stable origin becomes unstable. Instead, two additional fixed point attractors are born to either side of the trivial fixed point. This shift can be seen in Figure \ref{DufVanDer_alpha}(b). At $\alpha = -1.0$, another notable change is observed. The stochastic attractor associated with the LCO ceases to exist and the two fixed point attractors become dominant as shown in Fig. \ref{DufVanDer_alpha}(c).  A comparison of the PINN results with MC simulations demonstrates a high level of agreement.

\subsection{Example 2: The Duffing and Van Der Pol Oscillators under Additive and Multiplicative Noise}
In our second illustrative example, we consider two renowned oscillators under the combined effect of additive and multiplicative noise: the Duffing oscillator and the Van Der Pol oscillator. The joint PDF for such systems cannot be obtained by solving the associated FP equation analytically, or semi-analytically.  Thus, researchers often resort to numerical techniques. In this context, the PINN offers an alternative procedure to other classical numerical methods. 

In its basic form, the Duffing equation subject to additive and multiplicative white Gaussian noise can be expressed as:
\begin{align}
    &\Dot{x}_1 = x_2. \nonumber \\
    &\Ddot{x}_2 = -2\zeta\Dot{x}_2 - \alpha x_1 (1 + \eta_2 - \gamma^2 x_1^2) + \eta_1.
\end{align}
Following the formulation in Section \ref{sec:FP}, the corresponding stationary probability density $P(x_1, x_2)$ of the system is governed by the following FP equation: 
\begin{equation}
    0 = -\frac{\partial(x_2 P)}{\partial x_1} - \frac{\partial \left[ (\alpha x_1 (1 - \gamma^2 x_1^2) - 2 \zeta x_2 ) P\right]}{\partial x_2} + \frac{1}{2} \frac{\partial^2 [(S_{22} \alpha^2 x_1^2 + 2 S_{12} \alpha x_1 + S_{11}) P]}{\partial x_2^2}.
\end{equation}

\begin{figure}[t]
\centering
\includegraphics[width=0.85\textwidth]{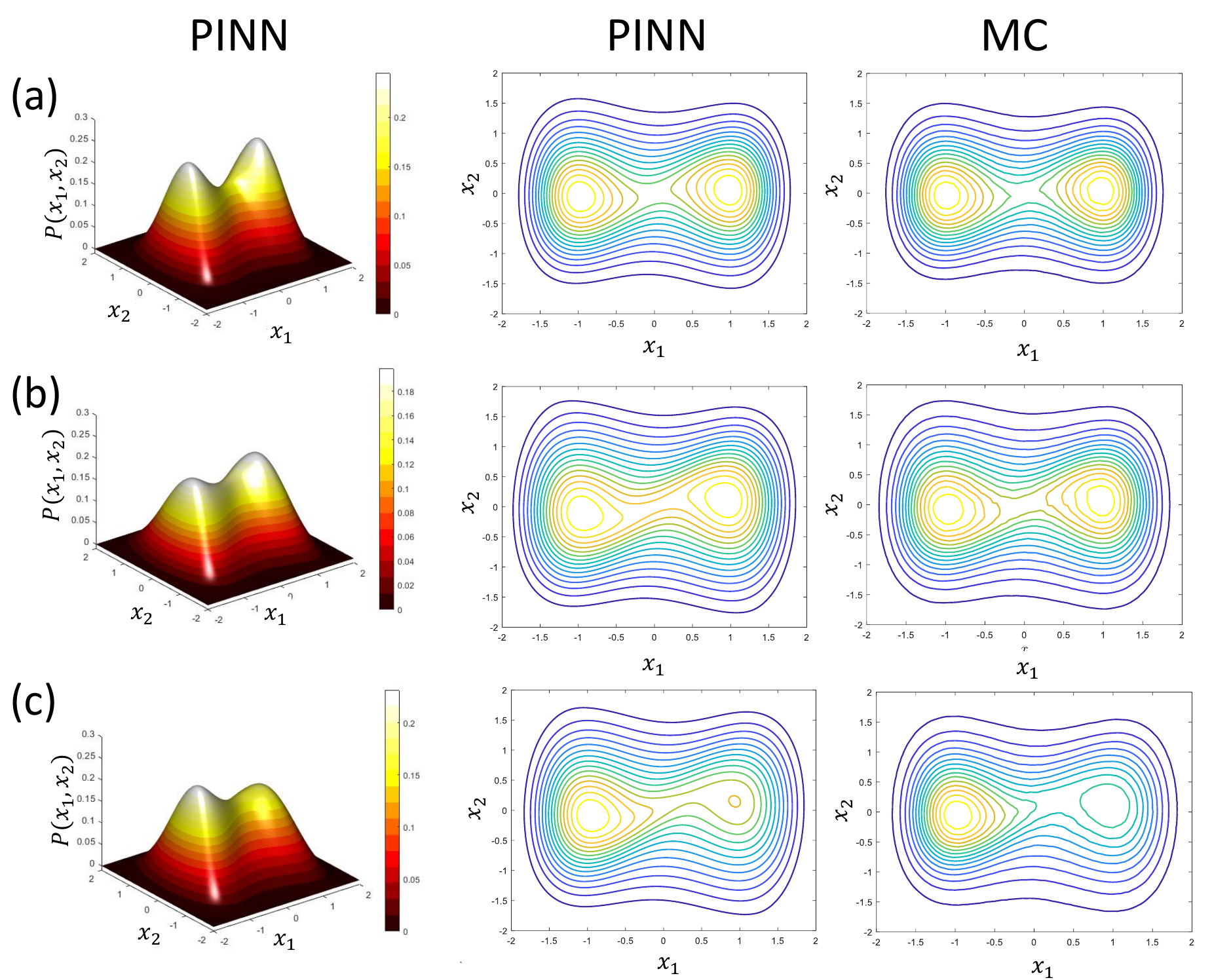}
\caption{Joint PDF of the Duffing oscillator under additive and multiplicative noise. Results are obtained for  $\zeta = 0.2, \alpha =-1, \gamma  = 1, S_{11}=0.05$, (a) $S_{12} =0,\; S_{22} =0.025 $, (b) $S_{12} =0,\; S_{22} =0.05 $, (c) $S_{12} =0.025,\; S_{22} =0.05$.}\label{fig:DuffingAddiMulti}.
\end{figure}

Results of the PINN prediction are demonstrated in Fig. \ref{fig:DuffingAddiMulti}. The figure deals with three distinct parameter sets. For each set, we generate the 3-dimensional surface representation of joint stationary PDF obtained by the PINN, and contour diagrams obtained using  PINN and the MC simulations, respectively. The system parameters and the additive noise remain the same in all cases, (i.e. $\zeta = 0.2, \alpha = -1, \gamma = 1, S_{11}=0.05$), while the multiplicative noise components, $S_{12}$ and $S_{22}$, are varied. Since $\alpha<0$, while $\gamma>0$, the potential energy function of the internal dynamics is bistable with two equilibrium states separated by a potential barrier at the origin.

For small values of the uncorrolated multiplicative noise $S_{22}=0.025$, the dynamic trajectories of the stochastic system has a high probability of being confined to a single potential well. Thus, two clear peaks appear in the joint PDF as shown in Fig. \ref{fig:DuffingAddiMulti}(a).  When the uncorrolated multiplicative noise increases to $0.05$, the peaks in the PDF become less pronounced and the probability of inter-well trajectories increases, Figs. \ref{fig:DuffingAddiMulti}(b). 

Increasing the positive cross-correlation between the multiplicative and additive noise, $S_{12} = 0.025$, as seen in Fig. \ref{fig:DuffingAddiMulti}(c), causes the joint PDF to lose its symmetry. The positive correlation has a tendency to rotate the PDF counter clockwise around the vertical axis. 
These predictions align closely with the MC simulations and  prior studies in the literature \cite{naprstek2008some, zeng2016impact}, further validating the robustness of the PINN framework in predicting the joint PDF under combined noise sources.

Next, we consider the stochastic response of the Van Der Pol oscillator under the combined effect of an additive noise term and a multiplicative noise appearing in the damping term. The equation of motion for such a system can be written as: 
\begin{align}
    &\Dot{x}_1 = x_2, \nonumber \\
    &\Dot{x}_2 = 2\zeta(1 + \eta_{2} - \beta_1 x_1^2)x_2 - \alpha x_1  + \eta_{1},
\end{align}
where $\beta_1>0$ is a nonlinear positive damping coefficient. Following the formulation in Equation \eqref{sec:FP}, the respective stationary FP equation reads: 
\begin{equation}
    0 = -\frac{\partial(x_2 P)}{\partial x_1} - \frac{\partial \left[ (\alpha x_1 - 2 \zeta (1- \beta_1^2 x_1^2 + \zeta S_{22}) - \zeta S_{12}) P\right]}{\partial x_2} + \frac{1}{2} \frac{\partial^2 [(4 S_{22} \zeta^2 x_2^2 + 4 S_{12} \zeta x_2 + S_{11}) P]}{\partial x_2^2}.
\end{equation}

\begin{figure}[t]
\centering
\includegraphics[width=0.85\textwidth]{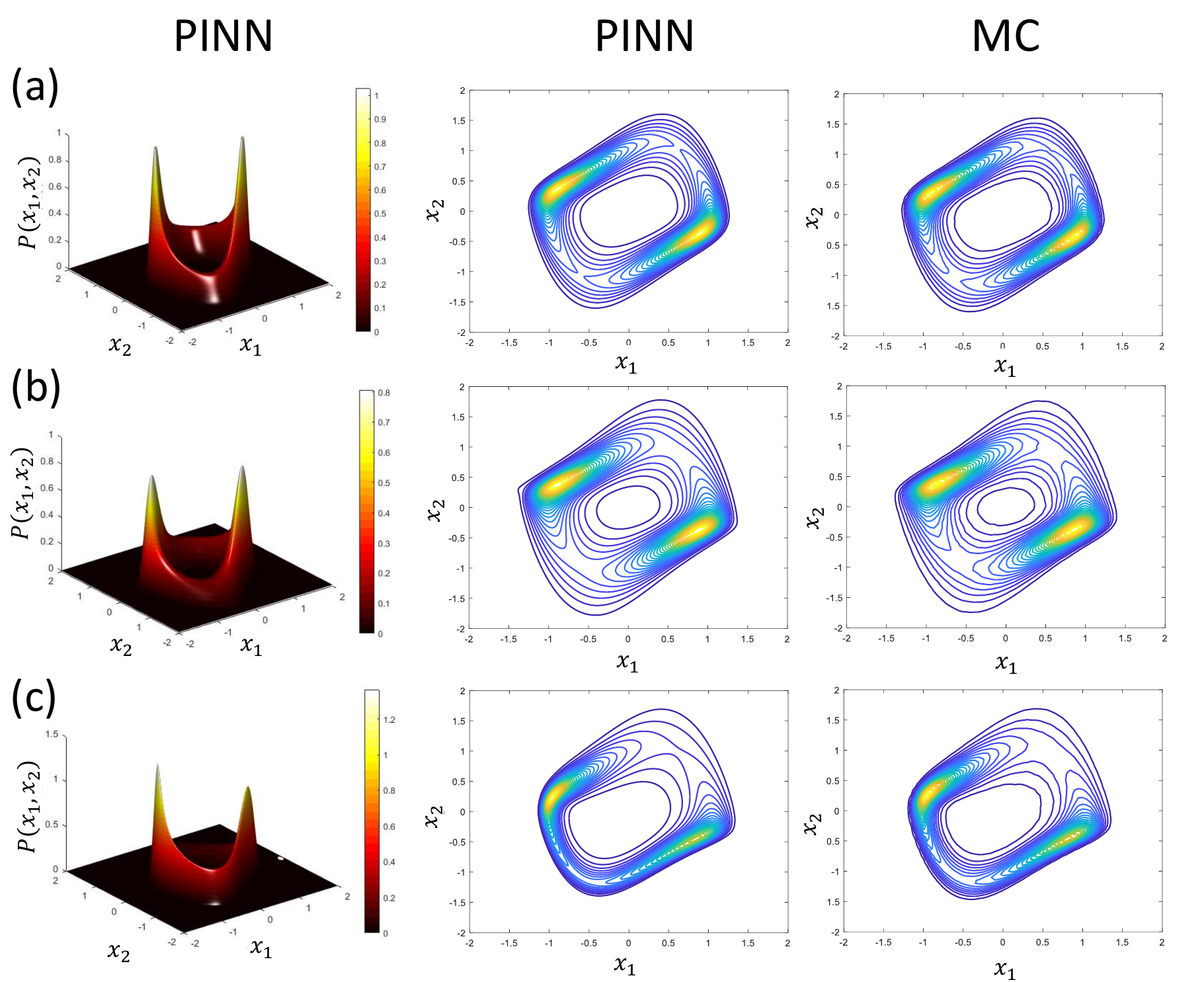}
\caption{Joint PDF of the Van Der Pol oscillator under additive and multiplicative noise. Results are obtained for  $\zeta = 0.3, \alpha =1, \beta =2, S_{11}=0.05$, (a) $S_{12} =0,\; S_{22} =0 $, (b) $S_{12} =0,\; S_{22} =0.2 $, (c) $S_{12} =0.05,\; S_{22} =0.05$.}\label{fig:VanDerPolAddiMulti}.
\end{figure}

We demonstrate and compare three distinct cases. The results corresponding to the specified parameters $\zeta = 0.3, \alpha =1, \beta_1 = 2, S_{11}= 0.05$ are shown in Fig. \ref{fig:VanDerPolAddiMulti}. The layout of the figure is the same as in Fig. \ref{fig:DuffingAddiMulti} for the Duffing system. It is clearly evident that the PINN can predict the topological variations of the PDF in terms of its shape and symmetry as the cross-correlation between the noise  terms changes. 

\subsection{Example 3: Duffing Oscillator under Band-Limited Noise}
In the preceding examples, we have explored cases under the influence of white noise (wide-band). While white noise is a valuable theoretical concept, in practice, it cannot exist because it has infinite energy. Alternatively, the more practical concept of band-limited noise can be created by passing white noise through a bypass filter \cite{to2000nonlinear}. This mathematical construct, however, increases the state-space of the FP equation. For band-limited noise, where the filter is a linear second-order ordinary differential equation \cite{zhu1993stochastic}, the resulting FP equation involves 4 independent state variables. 

In such cases, semi-analytical methods cannot be used to find the joint PDF. Computational methods have also been implemented but with limited success \cite{crandall1980non, er2011new}. Traditional computational methods are applicable only for a special class of problems, and extending them to fourth-order systems poses computational challenges as the number of unknowns in the computational domain are of the order of several millions \cite{kumar2014finite}. This places hard constraints on the usefulness of the computational methods for solving the FP equation for general high-order systems. As a result, in this context, PINNs may offer a powerful alternative. 

To illustrate this, we consider a nonlinear Duffing oscillator subjected to band-limited noise whose equation of motion can be written as
\begin{align}
    &\Dot{x}_1 = x_2, \nonumber \\
    &\Dot{x}_2 = - 2\zeta\,\Dot{x}_2 - \alpha\,x_1 - \gamma\,x_1^3 + y_1(t).
    \label{narrowbandosc}
\end{align}
Here, $y_1(t)$ is a band-limited stationary Gaussian process, which can be attained by passing white Gaussian noise through a bypass filter as following:
\begin{align}
    &\Dot{y}_1 = y_2, \nonumber \\
    &\Dot{y}_2 = -\delta \Dot{y}_2 - v^2 y_1 + v\sqrt{\delta} \eta_1,
    \label{narrowBandfilter}
\end{align}
where $\delta$ and $v$ are the bandwidth and the central frequency of the filter, respectively, and $\eta_1$ is a Gaussian white noise with a constant density $S_{11}$.  The joint stationary PDF, $P(x_1, x_2, y_1, y_2)$, is the solution of the following reduced FP equation: 
\begin{equation}   
0 = -x_2 \frac{\partial P}{\partial x_1} + \frac{\partial [(2\zeta x_2 + \alpha x_1 + \gamma x_1^3 - y_1)P]}{\partial x_2} - y_2 \frac{\partial P}{\partial y_1} + \frac{\partial ((\delta y_2 + v^2 y_1)P)}{\partial y_2} + \frac{1}{2}\delta v^2 S_{11} \frac{\partial^2 P}{\partial y_2^2}.
\end{equation}

We examine a nonlinear monostable system characterized by positive damping, linear, and cubic stiffness coefficients, $\alpha = 1$, $\gamma = 0.3$, and $\zeta = 0.1$. Additionally, the filter parameters are chosen such that $v = 1,\;\delta = 1$. We conduct a parametric analysis to determine the optimal network architecture and the number of collocation points required for accurate solutions in the four-dimensional case. Our findings indicate that a network comprising of 8 layers, 20 neurons, and a spacing of $h = 0.1$ yields accurate results. This outcome is particularly significant as it showcases the curse associated with adding higher dimension is less pronounced when compared with other classical solvers, such as FEM. To provide a clearer comparison,  this four-dimensional problem required only 230,000 points to solve, whereas in the FEM scenario, several million points are typically required to effectively tackle a similar four-dimensional problem  \cite{pichler2013numerical}. As such, considerable reduction of memory storage requirements is expected when using the PINN framework. 

\begin{figure}[t]
\centering
\includegraphics[width=0.7\textwidth]{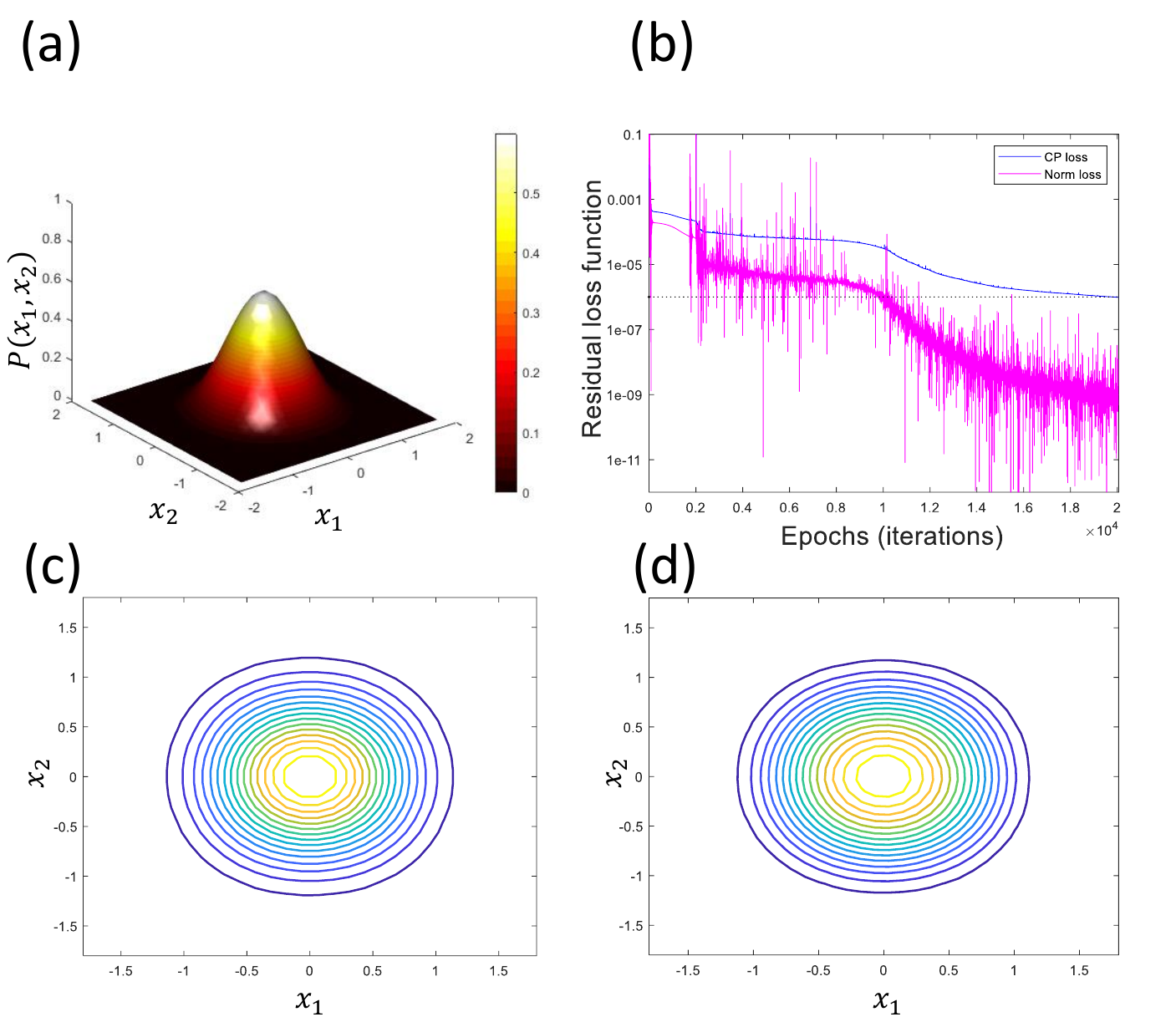}
\caption{(a) Marginal PDF surface obtained from the PINN framework. (b) Convergence of the residual loss functions, $\mathcal{L}_{PDE}$ and $\mathcal{L}_{norm}$. (c) Marginal PDF contour plot obtained from the PINN framework. (d) marginal PDF contour plot obtained from Monte-Carlo simulations. Results are obtained for $\zeta = 0.1$, $\alpha = 1$, $\gamma = 0.3$, $v = 1$, $\delta = 0.2$.}\label{fig:narrowBandCase}.
\end{figure}

Results are presented in Fig. \ref{fig:narrowBandCase}. Figure \ref{fig:narrowBandCase}(a) shows the marginal PDF obtained by integrating the joint PDF over the entire range of $y_1$ and $y_2$. Figure \ref{fig:narrowBandCase}(b) illustrates the convergence behavior of the loss terms, namely, the $\mathcal{L}_{PDE}$ and $\mathcal{L}_{norm}$. Both of these loss terms decrease to levels below $10^{-6}$ and $10^{-10}$, respectively. This behavior serves as a strong evidence for the successful fulfillment of both the PDE and the normalization constraints during the training process. Moreover, the reliability of the results can be reinforced by a qualitative comparison of the contour plots between the MC and PINN, as depicted in Figs. \ref{fig:narrowBandCase}(c) and (d). These plots reveal a good agreement between the two approaches. The PINN training converged after approximately 20,000 training epochs, and consumed around 30 minutes of computational time.  It is important to note that MC inherently carries approximations and is notably less accurate. Therefore, further quantitative evidence of the accuracy of the PINN will be shown next. 

We have conducted a quantitative numerical verification of the PINN approach. To achieve this, stationary samples of the narrow-band input, $y_1(t)$, were simulated by solving Equation \eqref{narrowBandfilter} with independently generated ensemble of white noise process samples. The resulting realizations of $y_1(t)$ were used as an input to Equation \eqref{narrowbandosc} for obtaining $x_1(t)$ and $x_2(t)$. Subsequently, we computed the variance of each sample after the transients had died out, and then averaged over the whole ensemble. Table \ref{comparison} shows a comparison between the variances obtained by the PINN approach and the numerical integration for different values of $r$. The ratio $r$ represents the ratio between the noise bandwidth and the bandwidth of the degenerate linear oscillator, $r = \delta/(2\zeta)$. It is observed that the PINN results and the numerical integration are in excellent agreement for all values of $r$, showing the PINN is a reliable technique to solve the PF solution even when the band-limited noise has a very narrow bandwidth as compared to the bandwidth of the oscillator.
\begin{table}[t]
\centering
\caption{Comparison of variances between the PINN approach and the numerical integration for different noise bandwidth ratios, $r$.}
\begin{tabular}{@{}lllll@{}}
\toprule
\multicolumn{1}{c}{$r$} & \multicolumn{2}{c}{PINN} & \multicolumn{2}{c}{\begin{tabular}[c]{@{}c@{}}Numerical \\ Integration\end{tabular}} \\ \midrule
                             & $\sigma^2_{x_1}$  & $\sigma^2_{x_2}$  & $\sigma^2_{x_1}$                                 & $\sigma^2_{x_2}$                                \\
2.5                         & 0.147       & 0.155      & 0.143                                     & 0.150                                    \\
1.0                         & 0.220       & 0.239      & 0.217                                     & 0.235                                    \\
0.5                          & 0.248       & 0.263      & 0.247                                     & 0.265                                    \\
0.25                         & 0.260       & 0.273      & 0.264                                     & 0.276                                    \\ \bottomrule
\end{tabular}
\label{comparison}
\end{table}

Even though the computational time required for the PINN is less than that of the MC, it is still relatively large. On average, the simulations required more than 30 minutes to converge. Here, we investigate improvements in the computational time by utilizing transfer learning. Transfer learning involves leveraging knowledge gained from a previously learned task to improve performance on a related, but slightly different task \cite{torrey2010transfer}. In this context, one can use the optimal $\theta$ values of a pre-trained model for a certain value of $r$ as initial guesses to predict the PDF for another value of $r$.  In this example, the pre-trained model pertains to $r = 0.5$, while we aim to adapt it for a new, previously unknown model with $r = 0.25$. 

\begin{figure}[t]
\centering
\includegraphics[width=0.7\textwidth]{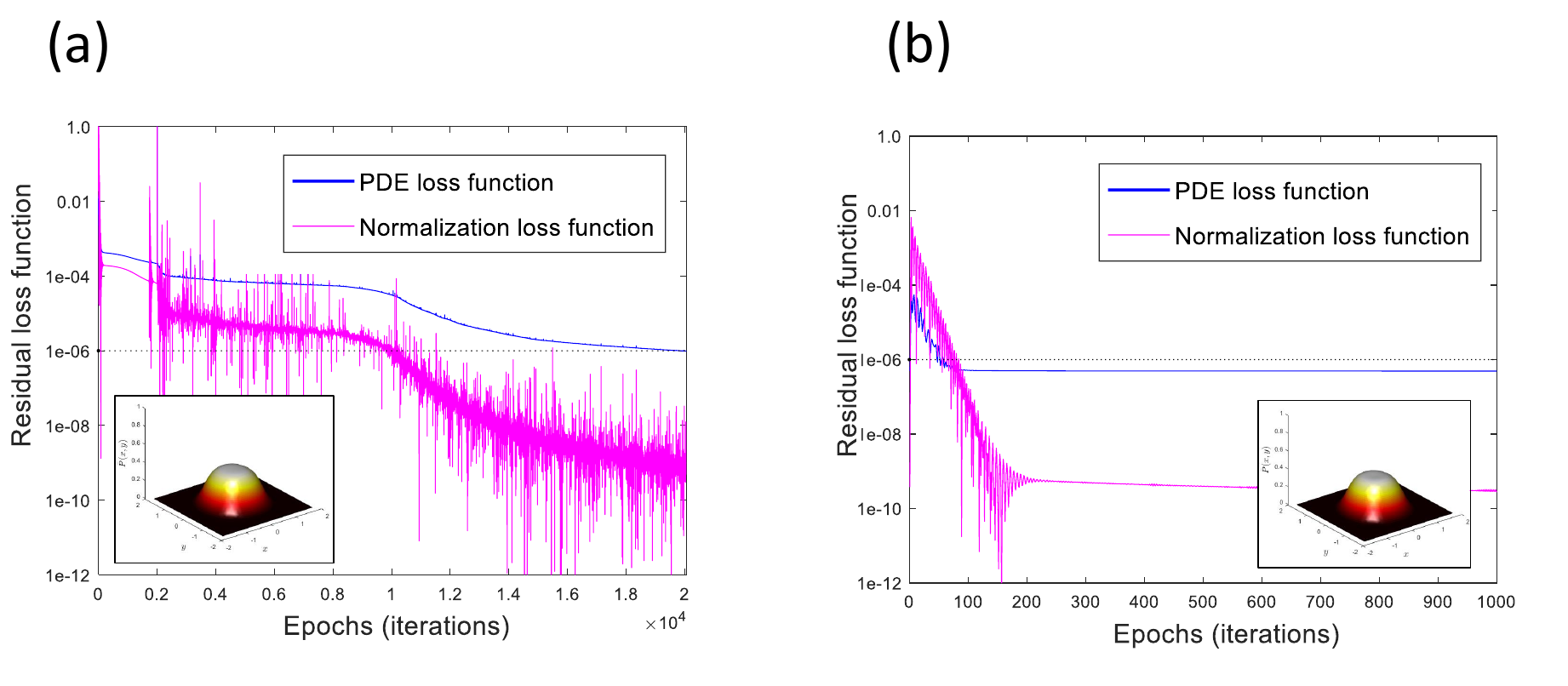}
\caption{Comparison of residual loss functions of the nonlinear Duffing oscillator under narrow-band noise with a bandwidth ratio $r= 0.25$, (a) random initialization, and (b) transfer learning from a pre-trained model of $r = 0.5$. Results are obtained for $\zeta = 0.1$, $\alpha = 1$, $\gamma = 0.3$, $v = 1$. }\label{fig:transferLearn}.
\end{figure}

A comparison of the residual loss functions of PINN  at $r = 0.25$ considering random initialization and those based on a pre-trained model is depicted in Figs. \ref{fig:transferLearn}(a) and (b), respectively. The results reveal that the PINN initialized with transfer learning achieved convergence in less than 500 epochs, whereas the random initialization approach required a substantially higher epochs to reach a similar level of convergence (epochs > 20000). This improvement of efficiency highlights the importance of transfer learning in the PINN framework especially when the computational resources are limited.

\section{Conclusions}
We developed a data-free, physics-informed approach to solve the FP equation for broad class of nonlinear dynamical systems. The method is based on fully-connected feed-forward network that takes a full-batch coordinates of the spatial domain and predicts the approximate stationary joint PDF via unconstrained optimization process. Two loss functions guide the training, namely, the PDE loss and the normalization loss. The boundary conditions are satisfied by construction of the network.

The approach is first validated for a stochastic system whose stationary PDF can be obtained through an exact analytical solution of the FP equation. Following the validation, a systematic parametric study is conducted to determine the optimal network architecture and the minimum number of collocation points necessary for convergence. Finally, the predictive capacity of the PINN framework is illustrated through several examples. For validation, results were compared with MC simulations, and when possible cross-referenced with published work in the literature. The main conclusions are as follows:
\begin{itemize}
    \item The PINN solver can detect the topological shifts (P-bifurcations) that occur in the stationary joint PDF of a nonlinear Duffing-Van Der Pol oscillators as some parameters are changed. Such bifurcations cannot be easily captured using semi-analytical tools; e.g. stochastic averaging. 
    
    \item The PINN solver can predict the response accurately under the combined effect of additive and multiplicative noise. Substantial reduction of the memory requirements is expected due to the meshless nature of PINN.  

    \item The PINN solver can accurately and effectively handle FP equations with a larger number of independent variables. This is illustrated by studying the response of a Duffing oscillator under band-limited noise, which is governed by four first-order stochastic differential equations. For such a high-dimensional system, the absence of analytical, and semi-analytical solutions of the FP equations as well as the high computational cost of the traditional FE/FD solvers highlight the PINN framework as a valuable and effective alternative.  A comparison of PINN results against numerical integration of the equation showed an excellent agreement even for very small noise bandwidths. Upon utilizing transfer learning, we showed that a substantial reduction of the computational time can be achieved.
\end{itemize}


\bibliography{sample}
\section*{Acknowledgements}
This research did not receive any specific grant from funding agencies in the public, commercial, or not-for-profit sectors.


\section*{Competing interests}
The author(s) declare no competing interests.
\end{document}